\documentclass[journal]{IEEEtran}
\usepackage{cite}
\usepackage{graphicx}

\ifCLASSINFOpdf
\else
\fi
\usepackage{amsmath}
\usepackage{algorithmic}
\usepackage{algorithm}
\usepackage{url}

\usepackage{multirow}
\usepackage{hhline}
\usepackage{array}
\usepackage{amsmath}
\usepackage{amssymb}
\usepackage{bm}
\usepackage{color}

\begin{document}
%
\title{Global Pixel Transformers for Virtual Staining of Microscopy Images}
%
%
%

\author{Yi~Liu,
        Hao~Yuan,
        Zhengyang~Wang,
        and~Shuiwang~Ji,~\IEEEmembership{Senior Member,~IEEE}
\thanks{Yi Liu and Shuiwang Ji are with the Department
of Computer Science \& Engineering, Texas A\&M University, College Station, TX 77843, USA.
Email: \{yiliu, sji\}@tamu.edu.}
\thanks{Hao Yuan is with School of Electrical Engineering and Computer Science,
WashinGPTon State University, Pullman, WA 99164, USA. Email: hao.yuan@wsu.edu.}}

\maketitle

\begin{abstract}
Visualizing the details of different cellular structures is of great
importance to elucidate cellular functions. However, it is
challenging to obtain high quality images of different structures
directly due to complex cellular environments. Fluorescence
staining is a popular technique to label different structures but
has several drawbacks. In particular, label staining is time consuming and
may affect cell morphology, and simultaneous labels are inherently
limited. This raises the need of building computational models to learn
relationships between unlabeled microscopy images and labeled fluorescence images,
and to infer fluorescence labels of other microscopy images 
excluding the physical staining process.
We propose to develop a novel deep model for virtual staining of 
unlabeled microscopy images. We first propose a novel network
layer, known as the global pixel transformer layer, that fuses global
information from inputs effectively. The proposed global pixel transformer layer can generate outputs with
arbitrary dimensions, and can be employed for all the regular, down-sampling, and up-sampling operators.
We then incorporate our proposed global pixel
transformer layers and dense blocks to build an U-Net like network.
We believe such a design can promote feature reusing between layers.
In addition, we propose a multi-scale input strategy to encourage
networks to capture features at different scales. We conduct
evaluations across various fluorescence image prediction
tasks to demonstrate the effectiveness of our approach. Both
quantitative and qualitative results show that our method
outperforms the state-of-the-art approach significantly. It is also
shown that our proposed global pixel transformer layer is useful to
improve the fluorescence image prediction results.
\end{abstract}


%
\IEEEpeerreviewmaketitle

\section{Introduction}

Capturing and visualizing the details of different sub-cellular
structures is an important but challenging problem in cellular
biology~\cite{koho2016image,jo2019quantitative}. Detailed
information on the shapes and locations of cellular structures plays
an important role in investigating cellular
functions~\cite{held2010cellcognition,glory2007automated,chou2008cell}.
The widely used transmitted light microscopy can only provide low
contrast images, and it is difficult to study certain structures or
functional characteristics from such
images~\cite{bray2012workflow,buchser2014assay}. One popular
technique to overcome these limitations is fluorescence staining,
which labels different structures with dyes or dye-conjugated
antibodies~\cite{ounkomol2018label}. For example, cell nuclei can be
labeled and visualized after being stained by
DAPI~\cite{ounkomol2018label,christiansen2018silico}. However,
fluorescence staining is time consuming, especially when cell
structures are complex. In addition, due to the overlap of spectrum,
there is a limit on the number of fluorescence labels to be applied
simultaneously on the same microscopy
image~\cite{bastiaens1999fluorescence,wang2010image}. Furthermore,
labeling may interfere with regular physiological processes in live
cells, resulting in changes in cell
morphology~\cite{jo2019quantitative,ounkomol2018label}. These
limitations raise the need of advanced methods to label cellular
structures more effectively and efficiently.


\begin{figure*}[t]
\centering
\includegraphics[width=0.8\textwidth, height=7cm]{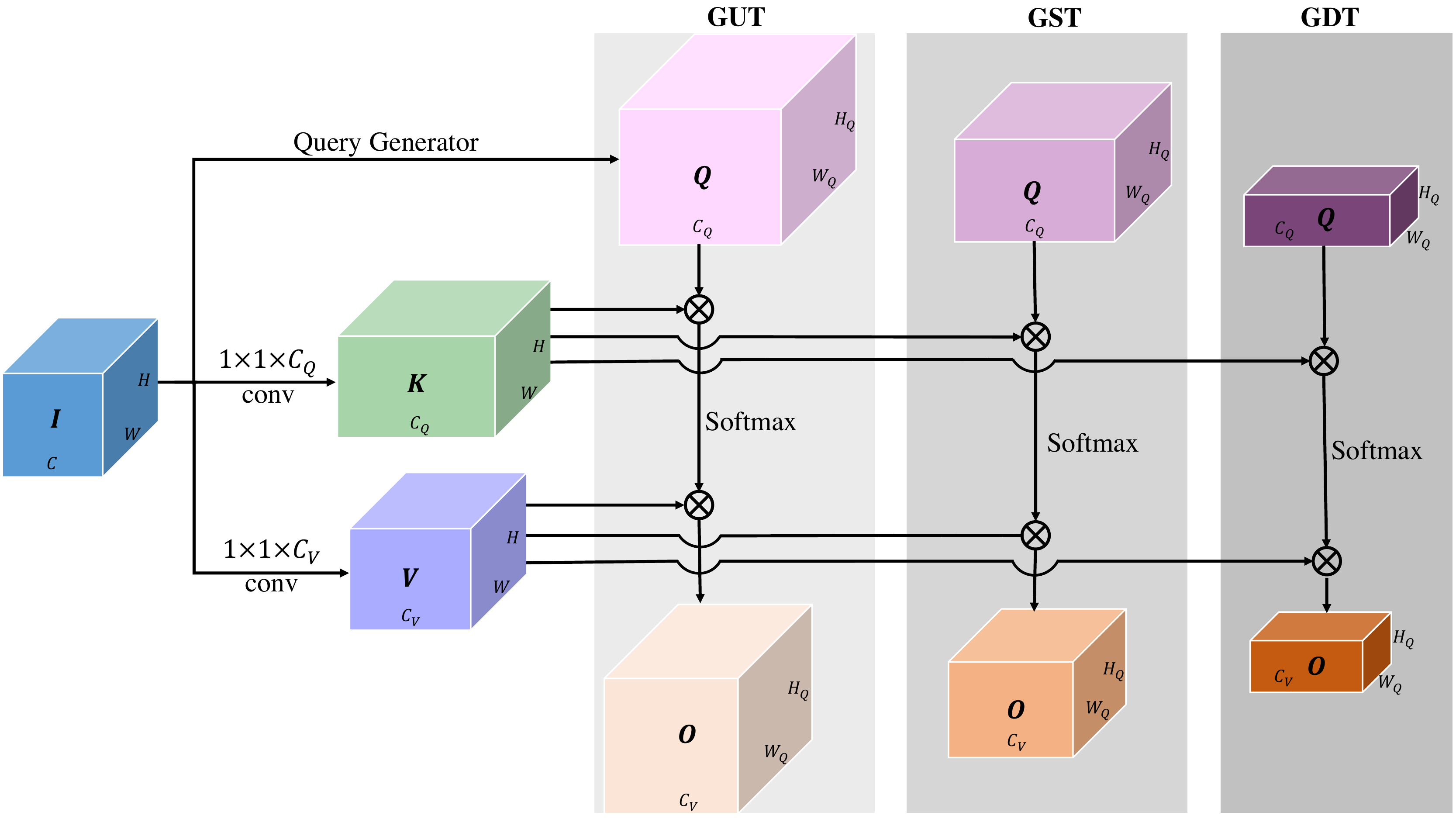}\vspace{-0.2cm}
\caption{Diagram of our proposed global pixel transformer (GPT) layer. The
spatial sizes of the output $\mathcal{O}$ is determined by $\mathcal{Q}$. Generally
speaking, a GPT layer can generate output $\mathcal{O}$ of arbitrary sizes. In
practice, three types of GPT layers are investigated. From left to right,
Global Up Transformer (GUT) layer doubles the spatial sizes; Global Same
Transformer (GST) layer keeps the spatial sizes; Global Down Transformer
(GDT) layer halves the spatial sizes. For each case, response at each
position in the output $\mathcal{O}$ is computed as a weighted summation of
features at all positions in $\mathcal{V}$, which is obtained directly from the
input features. Thus, global context information of the input $\mathcal{I}$ is
captured by a GPT layer.} \label{GPT_layer}
\vspace{-0.4 cm}
\end{figure*}

With the rapid development of deep learning methods, recent
studies~\cite{ounkomol2018label,christiansen2018silico,yuan2018computational}
propose to formulate such problems as image dense prediction tasks
using deep neural networks. In such a dense prediction task,  we
wish to predict if each pixel on the input microscopy image
belongs to a fluorescence label or not. Given microscopy
images and corresponding fluorescence stained images, the models are
trained to capture the relationship between them. Then for any newly
obtained microscopy image, the fluorescence image can be
predicted by the models based on the learned relationships.
Such a virtual staining process allows us to obtain 
fluorescence labels from microscopy
images without physical labeling~\cite{rivenson2019virtual}.

The recent study in~\cite{christiansen2018silico} proposes to use
convolutional neural networks
(CNNs)~\cite{lecun1998gradient,simonyan2014very,he2016deep} for such
a task and obtains promising results for prediction of fluorescence
images. It stacks multiple convolutional layers to enlarge the
receptive field and employs inception
modules~\cite{szegedy2015going} to facilitate
the training. However, only local operators, such as convolution,
pooling, and deconvolution, are used in their model. Hence, the
global information cannot be captured effectively and efficiently,
while such information may be important to determine certain
fluorescence labels. Meanwhile, another work~\cite{ounkomol2018label}
employs a vanilla U-Net framework for prediction of fluorescence
images. For each type of fluorescence label, it builds a model to
learn the relationships between microscopy images and the
corresponding fluorescence label. However, such a design learns
different fluorescence types separately, thereby ignoring important
relationships among different fluorescence labels. In addition, it
only employs local operators so that the global information cannot
be effectively captured. Other studies on fluorescence image super resolution~\cite{wang2019deep},
fluorescence image restoration~\cite{weigert2018content}, and image missing modality
prediction tasks~\cite{Cai:KDD18,Zhang:NI15,LiMICCAI14,Chen:KDD18} employ similar network
operators.

In this work, we propose a novel deep learning model, known as the
global pixel transformers (GPTs), for virtual staining of 
microscopy images.
As a radical departure from previous studies that invariably employ
local operators, we develop a novel network layer, known as the
global pixel transformer layer, to fuse global information efficiently and
effectively. The global pixel transformer layer is inspired by the
attention operators~\cite{vaswani2017attention,wang2018global}, and each position
of the output in the global pixel transformer layer fuses information from
all input positions. Particularly, our proposed layer can be flexibly
generalized to produce outputs of any dimensions. We build an U-Net
like architecture based on our proposed global pixel transformer layer. We
further develop dense blocks in our network to promote feature
reusing between layers in the network. To capture both global contextual
and local subtle features, we propose a multi-scale input strategy in our model to
incorporate information at different scales. particularly, our model
is designed in a multi-task manner to predict several target
fluorescence labels simultaneously. We conduct extensive
experiments to evaluate our proposed approach across various
fluorescence label prediction tasks. Both quantitative and
qualitative results show that our model outperforms the existing
approach~\cite{christiansen2018silico} significantly. Our ablation
analysis shows that the proposed global pixel transformer layer is useful
to improve model performance.






\begin{figure*}[t] \centering \includegraphics[width=0.9\textwidth]{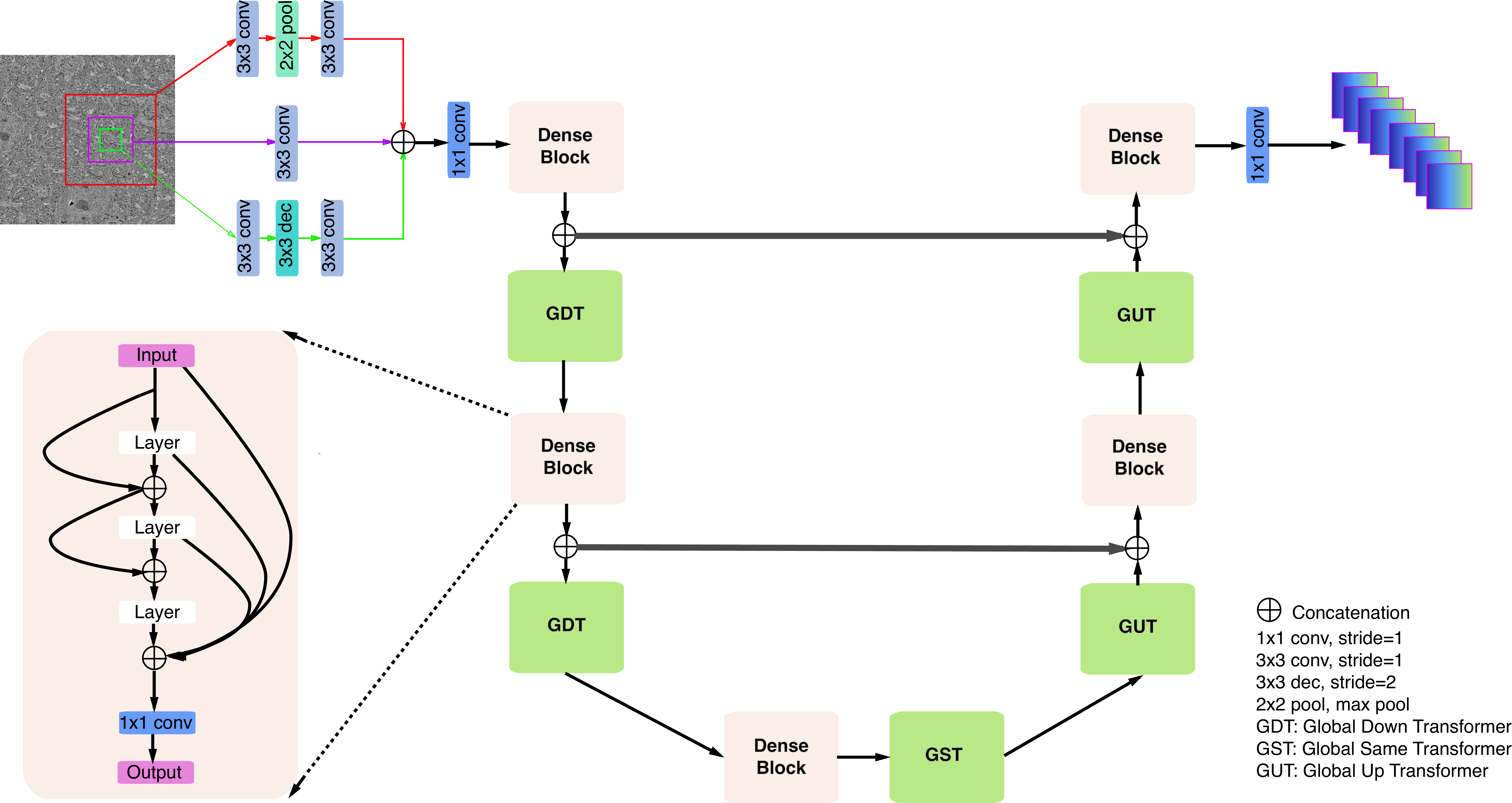}\vspace{-0.2cm}
\caption{Overall pipeline of our method for prediction of
fluorescence images. The network produces predications for a cropped
$H \times W$ patch in the whole image. For multi-scale inputs,
besides the cropped $H \times W$ patch, two other patches centered
at the same pixel with sizes $2H \times 2W$ and $H/2 \times W/2$ are
also cropped and re-scaled to $H \times W$. The input to the network
is the concatenation of these three patches. The U-like architecture
includes an encoder part and decoder part. In the encoder part, each
dense block is followed by a GDT layer. Sizes of feature maps are
reduced by the GDT layer, and numbers of feature maps are increased
by the dense block. In the decoder, each GUT layer is followed by a
dense block. GUT layers recover the spatial sizes and reduces the
number of feature maps. In the bottom block of the U-like
architecture, a GST layer following a dense block to transmit
information from the encoder to the decoder. A detailed diagram of a
dense block with 3 layers is also shown. Each layer includes
convolution, batch normalization, ReLU activation, and dropout.
A $1\times1$ convolution layer is added at the end
to adjust the number of feature maps.
Multi-task learning is used to predict several target fluorescence labels.} \label{Unet}
\end{figure*}

\section{Background and Related Work}
We describe the attention operator in this section.
The inputs to an attention operator include three matrices;
those are, a query matrix
$\bm{Q}= [\mathbf{q}_{1}, \mathbf{q}_{2}, \cdots, \mathbf{q}_{m}]\in\mathbb{R}^{c\times m}$
with each query vector $\mathbf{q}_{i}\in\mathbb{R}^c$, a key matrix
$\bm{K}= [\mathbf{k}_{1}, \mathbf{k}_{2}, \cdots, \mathbf{k}_{n}]\in\mathbb{R}^{c\times n}$
with each key vector $\mathbf{k}_{i}\in\mathbb{R}^c$, and a value matrix
$\bm{V}= [\mathbf{v}_{1}, \mathbf{v}_{2}, \cdots, \mathbf{v}_{n}]\in\mathbb{R}^{d\times n}$
with each value vector $\mathbf{v}_{i}\in\mathbb{R}^d$. An attention operator computes output
at each position by performing a weighted sum over all value vectors in $\bm{V}$, where the weights
are acquired by
attending the corresponding query vector
to all key
vectors in $\bm{K}$.
Formally, to compute a response at a position $i$, the attention operator
first computes the weight vector as
\begin{equation}\label{eq:att_weight_matrix}
\bm{a}_i = \mbox{Softmax}(\bm{K}^T\bm{q}_i)\in\mathbb{R}^{n},
\end{equation}
where $\mbox{Softmax}(\cdot)$ ensures the sum of all the elements in $\bm{a}_i$
to be 1.
Each element in $\bm{a}_i$
measures the importance of the corresponding vector in $\bm{K}$
by performing the inner product between it and $\mathbf{q}_{i}$.
The response at position $i$ is then
computed by using the weight vector $\bm{a}_i$ to
perform a weighted sum over all vectors in $\bm{V}$ as
\begin{equation}\label{eq:att_weight_sum}
\bm{o}_i = \bm{V}\bm{a}_i\in\mathbb{R}^d.
\end{equation}
In this way, the response at position $i$ fuses the global information
in $\bm{V}$ by assigning an importance to each value vector referring to $\mathbf{q}_{i}$.
For response at each position, we follow the same procedure and obtain outputs as
\begin{equation}\label{eq:final_o}
\bm{O} = [\bm{o}_{1}, \bm{o}_{2},...,\bm{o}_{m}]\in\mathbb{R}^{d\times m}.
\end{equation}
We rewrite outputs of an attention operator as
\begin{equation}\label{eq:total_o}
\bm{O} = \bm{V}\times\mbox{Softmax}(\bm{K}^T\bm{Q})\in\mathbb{R}^{d\times m},
\end{equation}
where $\mbox{Softmax}(\cdot)$ denotes a column-wise softmax operator to
ensure every column sum to 1. We can easily see the number of vectors in output matrix $\bm{O}$ is determined
by the number of vectors in query matrix $\bm{Q}$.
In self-attention operators, we set $\bm{Q}=\bm{K}=\bm{V}$.
Thus, response of a position is computed by the weighted average of features
at all positions, thereby fusing global information from input feature maps.
Note that a fully connected (FC) layer also fuses global information from whole
receptive fields. However, The self-attention operator computes responses based on
similarities between feature vectors at different positions, whereas a FC layer
connects every neuron to compute responses using learnable weights.
Moreover, a self-attention operator deals with inputs with variable sizes,
while an FC layer needs sizes of input to be fixed.

\section{Global Pixel Transformers} \label{sec:method}
In this section, we introduce a novel model for prediction of
fluorescence images, known as the multi-scale global pixel transformers with dense blocks.


\subsection{Global Pixel Transformer Layer}\label{GPT}

Traditional deep learning models for dense prediction tasks
contain several key operators,
such as convolution, pooling, and deconvolution. These operators
are all performed within a local neighborhood, restricting the capacity of
networks to fuse global
context information. To overcome this limitation, we propose a novel
network layer, known as the global pixel transformer (GPT) layer, which is based on
the attention operator and
captures dependencies between each position on outputs
and all positions on inputs,
thereby fusing global
information from input feature maps.
Unlike the self-attention operator that generates outputs with the same dimensions as the inputs,
our proposed GPT layer can generate output
feature maps with arbitrary dimensions, and can be employed for both
regular, down-sampling, and up-sampling operators. Specifically, we
investigate three types of global pixel transformer layers, namely global down
transformer (GDT) layer, global up transformer (GUT) layer, and global same
transformer (GST) layer. The dimensions of feature maps are halved in a GDT layer,
while those are doubled in a GUT layer and kept the same in a GST layer.



Although the three types of global pixel transformer layers generate outputs of
different sizes, they share similar structure and computational
pipeline. An illustration of our proposed GPT layer is provided in
Figure~\ref{GPT_layer}. Let $\mathcal{I}\in\mathbb{R}^{H\times W\times C}$ denote the
input of the GPT layer, the first step is to compute the query tensor $\mathcal{Q}$,
key tensor $\mathcal{K}$ and value tensor $\mathcal{V}$ based on $\mathcal{I}$. We employ a
generator layer to obtain the query tensor, and two $1\times1$
convolution layers to obtain the key and value tensors as
\begin{equation}\label{eq:QKV}
\begin{aligned}
\mathcal{Q} &= \mbox{Generator}(\mathcal{I})\in\mathbb{R}^{H_Q\times W_Q\times C_Q},\\
\mathcal{K} &= \mbox{Conv1}_{C_K}(\mathcal{I})\in\mathbb{R}^{H_K\times W_K\times C_K},\\
\mathcal{V} &= \mbox{Conv1}_{C_V}(\mathcal{I})\in\mathbb{R}^{H_V\times W_V\times C_V}, \\
\end{aligned}
\end{equation}
where Generator denotes a query generator layer, and $\mbox{Conv1}_M$ denotes a $1\times1$ convolution layer with
stride 1 and $M$ output feature maps. Hence, $H_K$ is equal to $H_V$
and $W_K$ is equal to $W_V$. The choice of the query generator
depends on the types of global pixel transformer layers. For GDT layers, we employ a
$3\times3$ convolutional layer with stride$=2$ to generate $\mathcal{Q}$. For
GUT layers, we employ a $3\times3$ deconvolutional layer with stride $=2$
to generate $\mathcal{Q}$. For GST layers, we employ a $3\times3$ convolutional layer
with stride$=1$ to generate $\mathcal{Q}$.

We then convert the each of the third-order tensors into a matrix by unfolding along mode-3~\cite{kolda2009tensor}.
In this way, tensor $\mathcal{Q}\in\mathbb{R}^{H_Q\times W_Q\times C_Q}$ is converted into a matrix
$\bm{Q}\in\mathbb{R}^{C_Q\times H_QW_Q}$. Similarly,
$\mathcal{K}\in\mathbb{R}^{H_K\times W_K\times C_K}$ is converted into a matrix
$\bm{K}\in\mathbb{R}^{C_K\times H_KW_K}$ and
$\mathcal{V}\in\mathbb{R}^{H_V\times W_V\times C_V}$ is converted into a matrix
$\bm{V}\in\mathbb{R}^{C_V\times H_VW_V}$. These three matrices serve as
the query, key and value matrices in Eq.~\ref{eq:total_o}.
To ensure the attention operator to be valid, we set ${C_K}={C_Q}$.
The output of the attention operator is computed as
\begin{equation}\label{eq:GPT_att}
\bm{O} = \bm{V}\times\mbox{Softmax}(\bm{K}^T\bm{Q})\in\mathbb{R}^{C_V\times H_QW_Q}.
\end{equation}



Finally, the output matrix $\bm{O}\in\mathbb{R}^{C_V\times H_QW_Q}$ is
converted back to a third-order tensor $\mathcal{O}\in\mathbb{R}^{H_Q\times W_Q\times C_V}$,
as output of the GPT layer. To this end,
each position feature in the output tensor $\mathcal{O}$ is computed as a weighted sum of all
feature vectors in $\mathcal{V}$, which
is obtained directly from the
input tensor $\mathcal{I}$. Apparently, global information from input features
is captured and fused to generate the output through our GPT layers.
In addition, the spatial sizes ($H_Q, W_Q$) of output feature maps
are determined by spatial sizes of the query tensor $\mathcal{Q}$, while the
number of output feature maps $C_V$ depends on the value tensor
$\mathcal{V}$. Theoretically, our proposed GPT layer can generate feature maps
of arbitrary dimensions. In practice, the commonly used local
operators either keep the spatial sizes of feature maps, or double
the spatial sizes for up-sampling, or halve the spatial sizes for
down-sampling. Hence, in this work, we propose to substitute these
local operators by three types of global pixel transformer layers.

The traditional local operators, such as $2\times 2$ max pooling
and convolution with a stride 2, may also capture global information
by stacking the same operator many times. However, such stacking is
not efficient. For example, when trying to capture the global
information in an $L \times L$ area, the $2\times 2$ max pooling
need to be repeated $ \lceil \log_2 L\rceil$ times. However, our
proposed GPT layers can capture global relationships among any two
positions using only one layer. Therefore, our proposed methods are
more efficient and effective compared to traditional local
operators.

\subsection{Global Pixel Transformers}\label{dunet}

It is well-known that encoder-decoder architectures like
U-Nets~\cite{ronneberger2015u} have achieved the state-of-the-art
performance in various dense prediction tasks. However, these
networks employ local operators like convolution, pooling and
deconvolution, which cannot efficiently capture global information.
Based on our GPT layer, we propose a novel network for dense
prediction tasks, known as the global pixel transformers~(GPTs).

In U-Nets, down-sampling layers are employed to reduce spatial sizes
and obtain high-level features, while up-sampling layers are used to
recover spatial dimensions. The commonly used convolution, pooling,
and deconvolution operators are performed in local neighborhood on
feature maps. We propose to substitute these local operators with
our proposed GPT layers. By setting different sizes for the query
tensor $Q$, our proposed GPT layers can be employed for both
down-sampling and up-sampling, while considering global information
to build output features. Suppose an input feature map has spatial
size of $H\times W$. For the down-sampling operator, a GDT layer
halves the spatial sizes of input feature maps, which can be
achieved by setting the sizes of query tensor as $H_Q=H/2$ and
$W_Q=W/2$. For the up-sampling operator, the spatial sizes of
feature maps are doubled by setting $H_Q=2H$ and $W_Q=2W$ in a GUT
layer. In addition, the GST layers are employed to transmit
information from the encoder to the decoder in the bottom block of
U-Nets.

In addition, due to the multiple down-sampling and up-sampling
operators in U-Nets, the spatial information, such as the shapes
and locations of cellular structures, is largely lost in its
information flow. Since the decoder recovers the spatial sizes from
high-level features, the prediction may not fully incorporate all
spatial information while such spatial information is important to
perform dense prediction. Hence, we adapt the idea to build skip
connections between the encoder and the decoder in U-Nets. Such
connections are expected to enable the sharing of spatial
information and high-level features between the encoder and decoder,
and hence improve the performance of dense prediction.

\begin{table*}[t]
\setlength{\tabcolsep}{5pt}
\renewcommand\arraystretch{1}
\begin{center}
\scriptsize \caption{A description of the datasets used in our
experiments. The datasets are created
by~\cite{christiansen2018silico} under five conditions from three
laboratories. A set of 13 2D images are z-stacks of
transmitted-light images collected from one 3D biomedical sample. In
total, eight fluorescence labels are introduced for all the
datasets.} \label{tb:data}
\begin{tabular}{c|llllccll}
\hline \bf {Condition} & \bf {Cell Type} & \textbf{\begin{tabular}[c]{@{}l@{}}fluorescence Label 1\\ and Modality\end{tabular}} & \textbf{\begin{tabular}[c]{@{}l@{}}fluorescence Label 2\\ and Modality\end{tabular}} & \textbf{\begin{tabular}[c]{@{}l@{}}fluorescence Label 3\\ and Modality\end{tabular}}
 & \textbf{\begin{tabular}[c]{@{}l@{}}Training\\ Data(2D)\end{tabular}}  & \textbf{\begin{tabular}[c]{@{}l@{}}Testing\\ Data(2D)\end{tabular}}  & \textbf{\begin{tabular}[c]{@{}l@{}}Spatial\\ Sizes\end{tabular}} & \bf {Laboratory}\\\hline
A & human motor neurons & DAPI (Wide Field) & TuJ1 (Wide Field) & Islet1 (Wide Field) & 286 & 39 & 1900x2600 & Rubin \\
B & human motor neurons & DAPI (Confocal) & MAP2 (Confocal) & NFH (Confocal) & 273 & 52 & 4600x4600 & Finkbeiner \\
C & primary rat cortical cultures & DAPI (Confocal) & DEAD (Confocal) & - & 936 & 273 & 2400x2400 & Finkbeiner \\
D & primary rat cortical cultures & DAPI (Confocal) & MAP2 (Confocal) & NFH (Confocal) & 26 & 13 & 4600x4600 & Finkbeiner \\
E & human breast cancer line & DAPI (Confocal) & CellMask (Confocal) & - & 13 & 13 & 3500x3500 & Google \\
\hline
\end{tabular}
\end{center}
\vspace{-10pt}
\end{table*}

\subsection{Global Pixel Transformers with Dense Blocks}\label{db}

To perform dense prediction on images, deep networks are usually
required to extract high-level features. However, a known problem
for training very deep CNNs is that gradient flow in deep networks
is sometimes saturated. Residual connections have been shown to be
effective to solve such a problem in various popular networks, such
as ResNets~\cite{he2016deep} and DenseNets~\cite{huang2017densely}.
In ResNets, residual connections are employed in residual blocks to
share the different levels of features between the non-linear
transformation of the input and the identity mapping. They benefit
the convergence of very deep neural networks by providing a highway
for the gradients to back propagate. Recently, residual
U-Net~\cite{quan2016fusionnet,fakhry2017residual} is proposed to
inherit the benefits of both long-range skip connections and
short-range residual connections. It is shown to obtain more precise
results on dense prediction tasks without increasing parameters.
Since DenseNets employ extreme residual connections, also known as
dense connections, to build dense blocks and achieve
state-of-the-art performance on image classification tasks, we
follow a similar idea to use dense blocks in our proposed global
transformer U-Nets.

The general structure of our model is shown in Figure~\ref{Unet}. We
combine the dense block and the GPT layer to better incorporate dense
connections. For the encoder part, each dense block is followed by a
GDT layer, since the dense block retains the spatial sizes of the
input while the GDT layer performs down-sampling. The reduction of
spatial sizes is compensated by the growth in feature map number
generated by the dense block. Correspondingly, each GUT layer in the
decoder is followed by a dense block, and the GUT layer recovers
the spatial sizes and reduces the number of feature maps.

For each dense block in our model, residual connections are employed
to connect every layer and its subsequent layers. A typical
$L-$layer dense block can be defined as
\begin{equation}\label{eq:db}
\begin{aligned}
x_L = H_L([x_0, x_1,...,x_{L-2},x_{L-1}]),
\end{aligned}
\end{equation}
where $x_0$ is the input to the dense block, $x_{l\in[1,...,L]}$ is
the output of the $l^{th}$ layer, and $[...]$ represents the
concatenation operator. $H_l(\cdot)$ denotes a series of
operators, including convolution, batch normalization
(BN)~\cite{ioffe2015batch}, ReLU activation, and
dropout~\cite{srivastava2014dropout}. Each layer in a dense block
generates $k$ new feature maps and they are concatenated with
previously generated feature maps. Note that $k$ is also called the
growth rate of dense block. Hence, the output of the dense block
contains information regarding both the input feature maps $x_0$ and
$k\times L$ newly generated feature maps. A general illustration of
our employed dense block is shown in Figure~\ref{Unet}. Note that we
add a $1\times1$ convolution layer before the output to make the
dense block more flexible so that the number of output feature maps
can be controlled. Intuitively, a dense block encourages feature
reusing between layers. In
addition, compared with traditional networks of the same capacity,
it can significantly reduce the number of parameters since each
layer in dense block only contains $k$ new feature maps.

\subsection{Multi-Scale Input Strategy}\label{multi}

One training strategy for dense prediction tasks is to feed the
whole image as input and produce predictions for all input pixels.
However, such a strategy requires excessive memory on training
hardware. On modern hardware like GPUs, memory resource is always
limited. This data feeding strategy becomes inefficient for large
inputs, which is quite common for biological image processing tasks.
One common solution is to crop small patches from the original
image, and train the neural networks with these small image patches.
To predict the whole image, an overlap-tile strategy can be used to
allow continuous segmentation~\cite{ronneberger2015u}. However, such
a divide-and-conquer strategy imposes a natural constraint on
networks. When predicting small patches, only the local information
within these patches can be captured by the network, while the
global information is ignored. Furthermore, the information in local
subtle area may be ignored when the sizes of local area are
relatively small compared with the patch sizes. To overcome these
limitations, we propose a multi-scale input strategy to incorporate
sufficient global and local information to perform prediction.

Assuming that the sizes of image patches for network training are
$H\times W$. For a image patch, let $(x_i, y_i)$ denote the center
and an $H\times W$ image patch $X_0$ is cropped for training. To
incorporate global information, we crop another $2H\times 2W$ image
$X_1$ with the same center to provide larger receptive field. This
$2H\times 2W$ image is re-scaled to $H\times W$ but contains more
global information. This is particular useful when the original
image $X_0$ contains pixels lying on incomplete edges. In addition,
we crop another $H/2\times W/2$ image $X_2$ to capture local subtle
information. The image $X_2$ is also re-scaled to $H\times W$.
Compared with $X_0$, small subtle areas are up-scaled in $X_2$,
which encourages the networks to capture important details. Then we
concatenate $X_0$, $X_1$ and $X_2$ along the channel dimension and
use them as input of networks. For the corresponding
label of such input, we use the predicted image of $X_0$
as its label. Intuitively, we incorporate information at different
scales to make predictions for one particular area. Notably, we can
flexibly generalize such input strategy to multiple levels and
incorporate information at different scales. Our proposed
multi-scale input strategy is illustrated in the left part of
Figure~\ref{Unet}.

\begin{table}[b]
\setlength{\tabcolsep}{3pt}
\renewcommand\arraystretch{1}
\begin{center}
\scriptsize
\caption{Detailed architecture of the proposed model used in our
experiments. $c$ denotes the number of classes. DB denotes a dense
block.} \label{tb:arch}
\begin{tabular}{l|cccc|c|c|c|c|c|c|c||c|cc}
\hline
\hhline{--------------} & \bf {Layers} & \bf {Spatial Sizes} & \bf {Channels} \\
\hhline{--------------}\multirow{3}*{Multi-Scale Input} & Input & Multi-Scaling Sizes & $c$  \\
&Multi-Scale Preprocessing  & 128x128 & $3c$  \\
&1x1 Convolution  & 128x128 & 32 \\
\hhline{--------------}\multirow{3}*{Encoder} &  DB(2 layers) + GDT & 64x64 & 64  \\
&DB(4 layers) + GDT & 32x32 & 128 \\
&DB(8 layers) + GDT & 16x16 & 256 \\
\hhline{--------------} Bottom Block & DB(8 layers) + GST & 16x16 & 384 \\
\hhline{--------------}\multirow{3}*{Decoder} & GUT + DB(4 layers) & 32x32 & 288 \\
&GUT + DB(2 layers) & 64x64 & 165 \\
&GUT + DB(1 layers) & 128x128 & 90 \\
\hhline{--------------} Output &1x1 Convolution & 128x128 & $c$  \\
\hline
\end{tabular}
\end{center}
\vspace{-10pt}
\end{table}

\begin{table*}[!t]
\begin{center}
\caption{Comparisons of Pearson's correlations on three tasks. For
the purpose of fair comparisons, we calculate the Pearson's
correlations for the baseline and our model on the same randomly
sampled pixels. Each time we randomly sample one million pixels and
calculate the Pearson's correlations. The results are obtained by
repeating the calculations 30 times, and we report the average and
standard deviation.} \label{tb:p}
\small
\begin{tabular}{c|cccc|c|cc|c|c|c|c|ccc}
\hline
&\multicolumn{4}{c|}{Cell Nuclei} & \multirow{2}*{Cell Viability} & \multirow{2}*{Cell Type}\\

& {Condition A} & {Condition B} & {Condition C} & {Condition D} & & \\
\hhline{--------------}Baseline & $0.928 \pm 0.0036$ & $0.871 \pm 0.0029$ & $0.920 \pm 0.0018$ & $0.902 \pm 0.0032$ & $0.852 \pm 0.0025$ & $0.839 \pm 0.0028$ \\
\hhline{--------------}Ours & \textbf{0.948 $\pm$ 0.0027}
& $\textbf{0.896 $\pm$ 0.0019}$ &
$\textbf{0.944 $\pm$ 0.0033}$ & $\textbf{0.915 $\pm$ 0.0031}$ &
$\textbf{0.859 $\pm$ 0.0022}$ & $\textbf{0.860 $\pm$ 0.0026}$ \\
\hline
\end{tabular}
\end{center}
\end{table*}

\section{Experimental Studies}

We use both quantitative and qualitative evaluations to demonstrate
the effectiveness of our proposed model. The dataset used for
evaluation and the experimental settings are presented in
Sections~\ref{data} and~\ref{setup}. We compare our experimental
results with the existing approach~\cite{christiansen2018silico} in
Section~\ref{res}. Finally, we provide an ablation analysis in
Section~\ref{ablation}.

\subsection{Dataset}~\label{data}

We use the dataset in the existing
work~\cite{christiansen2018silico}. The dataset contains 2D
high-resolution microscopy images from five different laboratories.
Note that a set of several such 2D microscopy images are originally
z-stacks of transmitted-light images collected from one 3D
biological sample~\cite{christiansen2018silico}. Specifically, the
z-stack 2D images are collected from several planes at equidistant
intervals along the z axis of a 3D sample. They collected 13
2D images from a sample. Thus, for all the 13 2D
images from the same set, they share the same fluorescence image
for each fluorescence label.
Different laboratories obtained the microscopy images under
different conditions using different methods. Two imaging
modalities, namely confocal and wide field are used during
microscopy photoing. In addition, three different types of cells are
collected by different laboratories, including human motor neurons
from induced pluripotent stem cells (iPSCs), primary rat cortical
cultures, and human breast cancer line. Detailed information of this
dataset is given in Table~\ref{tb:data}.

\begin{figure*}[!th]
\centering
\includegraphics[width=0.9\textwidth]{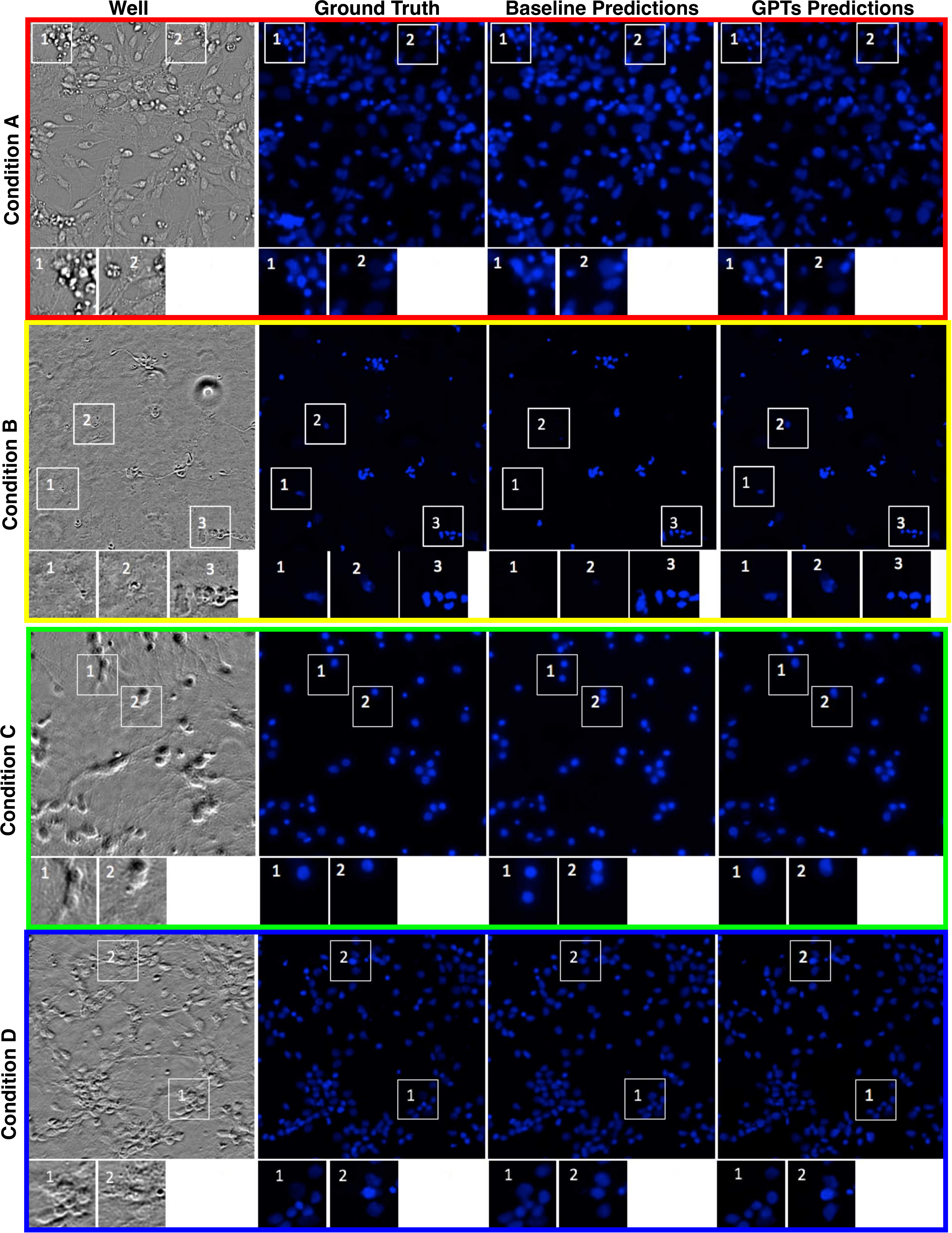}
\vspace{-0.2cm}
\caption{Visualization of prediction results for cell nuclei, which
are shown in blue. The first column is randomly cropped test
microscopy images from the datasets in Table~\ref{tb:data}. The
second column is the true fluorescence images for cell nuclei. The
third and fourth columns are predicted fluorescence images produced
by the baseline and our model, respectively.} \label{fig:nuclei}
\end{figure*}

\begin{figure*}[!tb]
\centering
\includegraphics[width=0.9\textwidth]{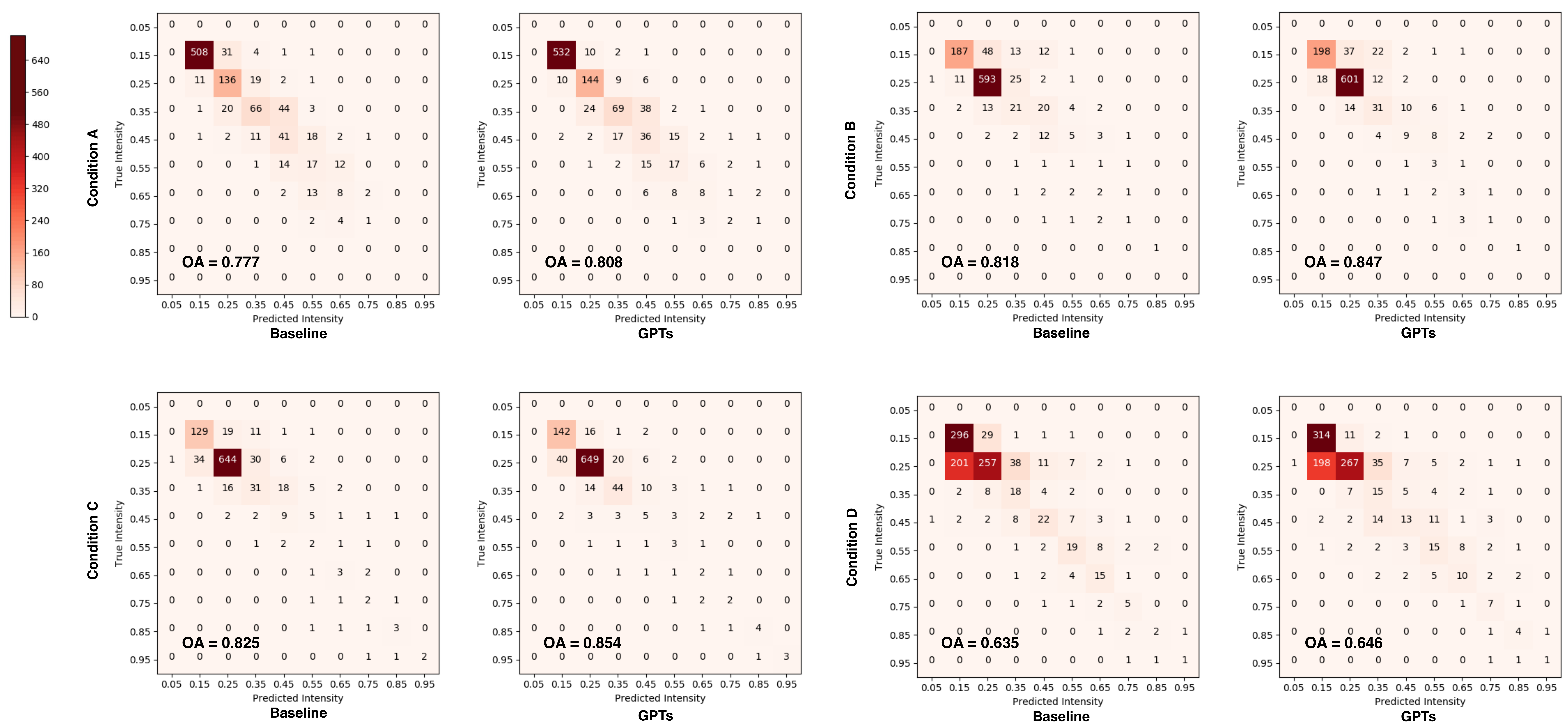}
\vspace{-0.2cm}
\caption{Confusion matrices based on the corresponding test images
in Figure~\ref{fig:nuclei}. Original pixel values in the range 0-255
are normalized to a scale of zero to one. The bin width is set to
0.1 on the normalized scale. The numbers in the bins are frequency
counts per 1,000.} \label{fig:confu1}
\end{figure*}

\begin{table*}[!t]
\begin{center}
\caption{Ablation analysis on prediction of cell nuclei by comparing
Pearson's correlations between different models. DB denotes dense
block. All models are trained across all training samples and
evaluated on one specific task. Details of the models are provided
in Section~\ref{sec:method}.}
\label{tb:aba}
\small
\begin{tabular}{l|cccc c|c|c|c|c|c|c||c|cc}
\hline
\hhline{--------------} &{Condition A} & {Condition B} & {Condition C} & {Condition D}\\
\hhline{--------------}{Baseline} & 0.928 & 0.871 & 0.920 & 0.902 \\
{Multi-scale U-Nets} & 0.937 & 0.882 & 0.925 & 0.893 \\
{Multi-scale U-Nets with DBs} & 0.941 & 0.887 & 0.935 & 0.902 \\
{Multi-scale GPTs with DBs} & 0.948 & 0.896 & 0.944 & 0.915 \\
\hline
\end{tabular}
\end{center}
\vspace{-10pt}
\end{table*}

\subsection{Experimental Setup}~\label{setup}

The architecture of our model is shown in Table~\ref{tb:arch}. It
shows the changes of feature maps through the information flow in
our networks. The growth rate of our dense blocks is set to 16.  We
employ three GDT layers with dense blocks in our encoder to perform
down-sampling and extract high-level features. Correspondingly,
there are three GUT layers with dense blocks to recover the spatial
sizes. For the bottom block connecting the encoder and the decoder,
we employ one GST layer and one dense block. Note that the depths of
different dense blocks are different.

\begin{figure*}[!tb]
\centering
\includegraphics[width=0.9\textwidth]{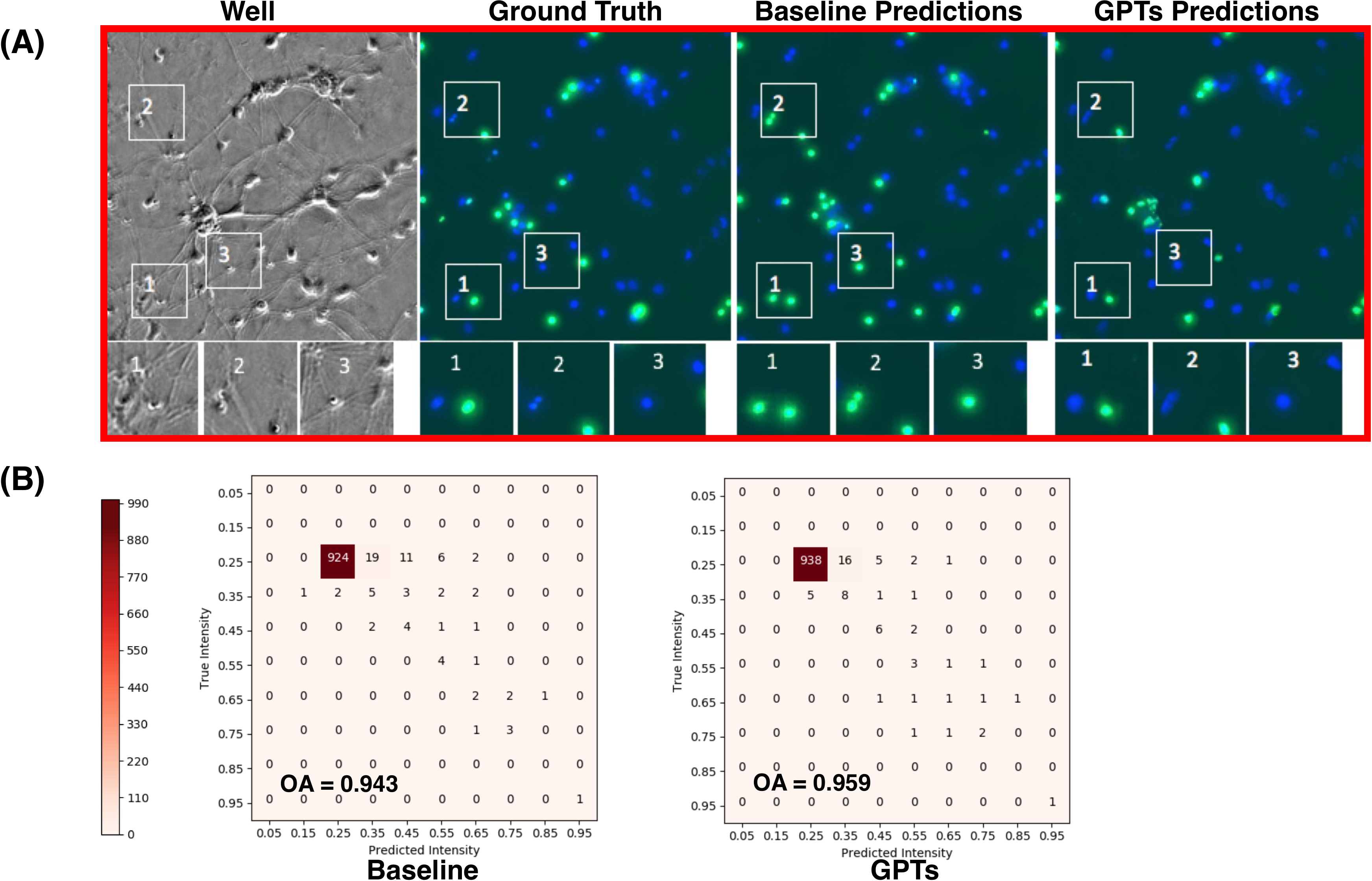}
\vspace{-0.2cm}
\caption{Visualization of prediction results for dead cells. (A) The
first column is randomly cropped test microscopy images on the
datasets from condition C in Table~\ref{tb:data}. The second column
is the true fluorescence images for dead cells. The third and fourth
columns are predicted fluorescence images produced by the baseline
and our model, respectively. For all the fluorescence images, dead
cells are shown in green. The corresponding cell nuclei are shown in
blue and added to the fluorescence images as visual background. (B)
Pixel values in the range 0-255 are normalized to zero to one. The
bin width is set to 0.1. The numbers in the bins are frequency
counts per 1,000.} \label{fig:via}
\vspace{-5pt}
\end{figure*}

Training examples are obtained by randomly cropping from the raw
images. Since we employ the multi-scale input strategy, we crop
images at three different scales; namely $64\times 64$, $128\times
128$, and $256\times 256$. The network predicts fluorescence maps
with sizes equal to $128\times 128$. We train our proposed model
across all training examples in a multi-task learning manner. Since
there are eight fluorescence labels, our model learns eight tasks
simultaneously to capture and refine common features across all
training samples. Specifically, for each input image, our model
generates eight $128\times 128$ fluorescence maps, and each map
corresponds to one fluorescence label. In addition, for each pixel in
the predicted maps, the network outputs a probability distribution
over 256 pixel values, so $c=256$ in Table~\ref{tb:arch}.
Cross-entropy loss is employed for network training. Note that there
are at most three fluorescence labels available for a given input.
The loss is calculated by only considering target labels while
irrelevant labels are ignored. During training, we employ the
dropout with a rate of 0.5 in our dense blocks to avoid
over-fitting. To optimize the model, we employ the Adam
optimizer~\cite{kingma2014adam} with a learning rate of $1\times
e^{-5}$ and a batch size of 4. During the prediction stage,
test patches are cropped in a
sliding-window fashion. We extract patches from test images with the
same sizes as those in training ($128\times 128$) by sliding a
window with a constant step size. The step size is set to 64 in our
experiments. Then we build predictions for the original test images
based on predictions of small patches.

\begin{figure*}[ht]
\centering
\includegraphics[width=0.9\textwidth]{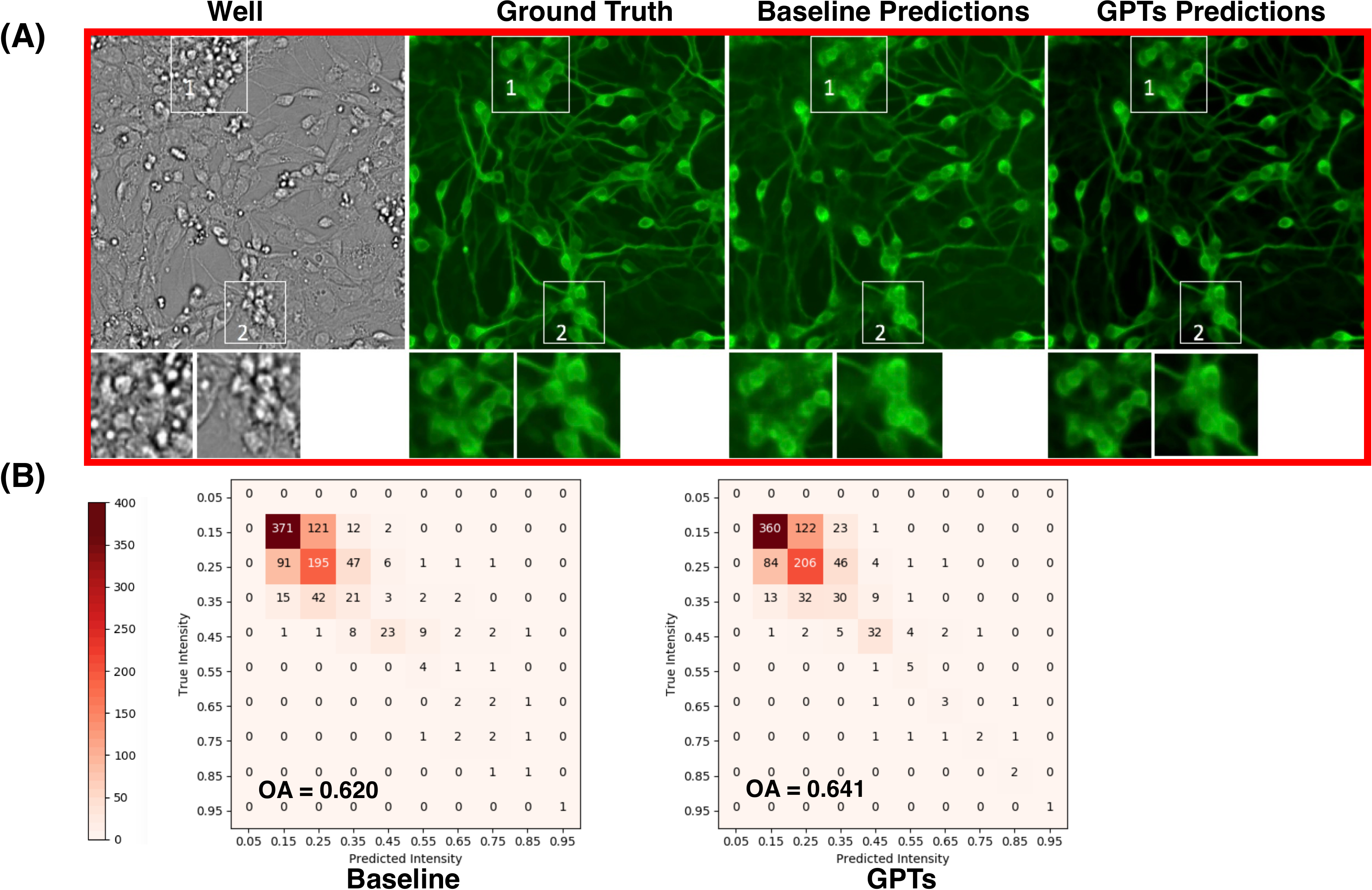}
\vspace{-0.2cm}
\caption{Visualization of prediction results for cell type as
neurons. (A) The first column is the randomly cropped test
microscopy images on the datasets from condition A in
Table~\ref{tb:data}. The second column is the true fluorescence
images for neurons. The third and fourth columns are predicted
fluorescence images produced by the baseline and our model,
respectively. For all the fluorescence images, neurons are shown in
green. (B) Original pixel values in the range 0-255 are
normalized to a scale of zero to one. The bin width is set to 0.1 on
the normalized scale. The numbers in the bins are frequency counts
per 1,000.} \label{fig:type}
\end{figure*}

\begin{table*}[!t]
\begin{center}
\small
\caption{Comparisons of accuracies for each bin and the overall
accuracies based on the confusion matrices for all the tasks. `-'
represents no true pixel values lie in this bin.} \label{tb:accu}
\begin{tabular}{l|cccccccccccc cc}
\hline
\hhline{--------------}Scenarios & &{Bin0} & {Bin1} & {Bin2} & {Bin3}&{Bin4} &{Bin5} &{Bin6} &{Bin7} &{Bin8} &{Bin9} &{Overall} \\
\hhline{--------------}\multirow{2}*{Cell Nuclei-Condition A} & Baseline & - & 0.932 & 0.805 & 0.493& 0.539& 0.386& 0.320& 0.143& -& -& 0.777\\
 &Ours & - & 0.976 & 0.852 & 0.515& 0.474& 0.386& 0.320& 0.286& -& -& \textbf{0.808}\\
 \hhline{--------------}\multirow{2}*{Cell Nuclei-Condition B} & Baseline & - & 0.716 & 0.937 & 0.339& 0.480& 0.200& 0.250& 0.200& 1.000& -& 0.818\\
&Ours & - & 0.759 & 0.950 & 0.500& 0.360& 0.600& 0.375& 0.200& 1.000& -& \textbf{0.847}\\
 \hhline{--------------}\multirow{2}*{Cell Nuclei-Condition C} & Baseline & - & 0.801 & 0.898 & 0.425& 0.429& 0.286& 0.500& 0.400& 0.500& 0.500& 0.825\\
 &Ours & - & 0.882 & 0.905 & 0.603& 0.238& 0.429& 0.333& 0.400& 0.667& 0.75& \textbf{0.854}\\
 \hhline{--------------}\multirow{2}*{Cell Nuclei-Condition D} & Baseline &-& 0.902 & 0.498 & 0.529 & 0.478& 0.559& 0.652& 0.556& 0.333& 0.333& 0.635\\
 &Ours & - & 0.957 & 0.516 & 0.441& 0.283& 0.441& 0.435& 0.778& 0.667& 0.333& \textbf{0.646}\\
 \hhline{--------------}\multirow{2}*{Cell Viability} & Baseline & - & - & 0.960 & 0.333& 0.500& 0.800& 0.400& 0.750& -& 1.000& 0.943\\
 &Ours &- & - & 0.975 & 0.533 & 0.750 & 0.600& 0.200& 0.500& -& 1.000& \textbf{0.959}\\
 \hhline{--------------}\multirow{2}*{Cell Type} & Baseline & - & 0.733 & 0.570 & 0.247& 0.489& 0.667& 0.400& 0.333& 0.500& 1.000& 0.620\\
 &Ours & - & 0.711 & 0.602 & 0.353& 0.681& 0.833& 0.600& 0.333& 1.000& 1.000& \textbf{0.641}\\
\hline
\end{tabular}
\end{center}
\vspace{-10pt}
\end{table*}

\subsection{Comparison with the Baseline}\label{res}
We compare our approach with the existing
model~\cite{christiansen2018silico} as it achieves the
state-of-the-art performance on the dataset we are using. To
demonstrate the effectiveness of our proposed approach, we conduct
comparisons with the baseline method for three different tasks:

\textbf{Prediction of Cell Nuclei}: Given an image, the task is to
predict the nuclei of live cells. The nuclei of live cells are
labeled using DAPI on both confocal and wild field modalities.
Examples created under condition A, B, C, D have fluorescence labels
to investigate the cell nuclei.

\textbf{Prediction of Cell Viability}: Given an image, this task
predicts the dead cells with cell nuclei as visual background.
Dead cells on images are labeled with
propidium lodide (PI) on confocal modality. These images are
obtained under condition C.

\textbf{Prediction of Cell Type}: Given an image, this task predicts
the neurons with cell nuclei as visual background.
There may exist two other types of cells in the image,
such as astrocytes and immature dividing cells. Neurons on images
are labeled using TuJ1 under condition A.

We first compare our approach with the baseline method
quantitatively, using  Pearson's correlation values calculated for
each task. Following the work~\cite{christiansen2018silico}, one million pixels are randomly sampled
from all the test images in a task, and we collect the predicted
values for these pixels. These predicted results can be represented
as a one million dimensional vector. Similarly, we can obtain
another one million dimensional vector from the ground truth of
these pixels. Then we calculate the Pearson's correlation between
these two vectors, which can indicate the similarity between them.
In particular, higher Pearson's correlation values imply that the
predicted results are closer to the ground truth. The results are
reported in Table~\ref{tb:p}. Note that for both our method and the
baseline approach, we repeat the calculations 30 times and report
the average and standard deviation. We can observe that the proposed
model outperforms the baseline model significantly on all of the
three tasks. These results indicate that the proposed model can
better capture the relationships between microscopy images
and the corresponding fluorescence labels.

In addition, we compare the prediction results qualitatively. We
present the prediction results for the cell nuclei task in
Figure~\ref{fig:nuclei}. Based on visual comparisons for the areas
in white boxes, we can observe that our model can make more accurate
predictions for many small regions. These results demonstrate the
capability of our model to capture detailed information.
Furthermore, confusion matrices are reported for these images to
allow visualization of true versus predicted pixel values in each
bin. The pixel values are normalized to $[0, 1]$ and divided into 10
bins that the $i^{th}$ bin contains the pixels with values in the
range $[0.1*i, 0.1+0.1*i)$. The overall accuracies (OAs) in
confusion matrices indicate how many pixels
are classified into the same bin as the ground truth. As shown in
Figure~\ref{fig:confu1}, our model can predict more accurate pixel
values compared with the baseline model. Similarly, we
report the prediction results and the corresponding confusion
matrices for the dead cell task in Figure~\ref{fig:via}. The white
boxes show that the baseline misclassifies dead cells to other
labels while our model has the ability to make correct predictions.
We also show the results of the cell type task in
Figure~\ref{fig:type}. We can clearly observe that our model
achieves more accurate predictions on neurons.
Finally, we report the prediction accuracies for different bins and the overall
accuracies in Table~\ref{tb:accu}. Obviously, for all three task, we obtain more accurate predictions.
Overall, both qualitative and quantitative
results indicate that our model performs significantly better than
the baseline approach.

\subsection{Ablation Analysis}\label{ablation}
We conduct ablation analysis on the cell nuclei prediction task to
show the effectiveness of each proposed module. All models are
trained under the same condition and compared with fair settings. As
shown in Table~\ref{tb:aba}, when employing the multi-scale input
strategy, even the classic U-Nets can achieve better results than
the baseline approach. By adapting to dense blocks, the performance
is further improved. The best performance is achieved by
incorporating all of our proposed modules. Such results indicate
that all of our proposed modules are effective to improve predictive
performance.

\section{Conclusion}
Visualizing cellular structure is important to understand cellular
functions. Fluorescence staining is a popular technique but has
key limitations. Here, we develop a novel deep learning model to
directly predict labeled fluorescence images from unlabeled microscopy images.
To fuse global information efficiently and effectively, we propose a
novel global pixel transformer layer and build an U-Net like network by
incorporating our proposed global pixel transformer layer and dense
blocks. A novel multi-scale input strategy is also proposed to
combine both global and local features for more accurate
predictions. Experimental results on various fluorescence image
prediction tasks indicates that our model outperforms the baseline
model significantly. In addition, ablation study shows that all of
our proposed modules are effective to improve performance.


%




\section*{Acknowledgment}
This work was supported by National Science Foundation
[IIS-1633359, IIS-1615035, and DBI-1641223].




\bibliography{deep}

\begin{thebibliography}{10}
\providecommand{\url}[1]{#1}
\csname url@samestyle\endcsname
\providecommand{\newblock}{\relax}
\providecommand{\bibinfo}[2]{#2}
\providecommand{\BIBentrySTDinterwordspacing}{\spaceskip=0pt\relax}
\providecommand{\BIBentryALTinterwordstretchfactor}{4}
\providecommand{\BIBentryALTinterwordspacing}{\spaceskip=\fontdimen2\font plus
\BIBentryALTinterwordstretchfactor\fontdimen3\font minus
  \fontdimen4\font\relax}
\providecommand{\BIBforeignlanguage}[2]{{%
\expandafter\ifx\csname l@#1\endcsname\relax
\typeout{** WARNING: IEEEtran.bst: No hyphenation pattern has been}%
\typeout{** loaded for the language `#1'. Using the pattern for}%
\typeout{** the default language instead.}%
\else
\language=\csname l@#1\endcsname
\fi
#2}}
\providecommand{\BIBdecl}{\relax}
\BIBdecl

\bibitem{koho2016image}
S.~Koho, E.~Fazeli, J.~E. Eriksson, and P.~E. H{\"a}nninen, ``Image quality
  ranking method for microscopy,'' \emph{Scientific reports}, vol.~6, p. 28962,
  2016.

\bibitem{jo2019quantitative}
Y.~Jo, H.~Cho, S.~Y. Lee, G.~Choi, G.~Kim, H.-s. Min, and Y.~Park,
  ``Quantitative phase imaging and artificial intelligence: A review,''
  \emph{IEEE Journal of Selected Topics in Quantum Electronics}, vol.~25,
  no.~1, pp. 1--14, 2019.

\bibitem{held2010cellcognition}
M.~Held, M.~H. Schmitz, B.~Fischer, T.~Walter, B.~Neumann, M.~H. Olma,
  M.~Peter, J.~Ellenberg, and D.~W. Gerlich, ``Cellcognition: time-resolved
  phenotype annotation in high-throughput live cell imaging,'' \emph{Nature
  methods}, vol.~7, no.~9, p. 747, 2010.

\bibitem{glory2007automated}
E.~Glory and R.~F. Murphy, ``Automated subcellular location determination and
  high-throughput microscopy,'' \emph{Developmental cell}, vol.~12, no.~1, pp.
  7--16, 2007.

\bibitem{chou2008cell}
K.-C. Chou and H.-B. Shen, ``Cell-ploc: a package of web servers for predicting
  subcellular localization of proteins in various organisms,'' \emph{Nature
  protocols}, vol.~3, no.~2, p. 153, 2008.

\bibitem{bray2012workflow}
M.-A. Bray, A.~N. Fraser, T.~P. Hasaka, and A.~E. Carpenter, ``Workflow and
  metrics for image quality control in large-scale high-content screens,''
  \emph{Journal of biomolecular screening}, vol.~17, no.~2, pp. 266--274, 2012.

\bibitem{buchser2014assay}
W.~Buchser, M.~Collins, T.~Garyantes, R.~Guha, S.~Haney, V.~Lemmon, Z.~Li, and
  O.~J. Trask, ``Assay development guidelines for image-based high content
  screening, high content analysis and high content imaging,'' 2014.

\bibitem{ounkomol2018label}
C.~Ounkomol, S.~Seshamani, M.~M. Maleckar, F.~Collman, and G.~R. Johnson,
  ``Label-free prediction of three-dimensional fluorescence images from
  transmitted-light microscopy,'' \emph{Nature methods}, vol.~15, no.~11, p.
  917, 2018.

\bibitem{christiansen2018silico}
E.~M. Christiansen, S.~J. Yang, D.~M. Ando, A.~Javaherian, G.~Skibinski,
  S.~Lipnick, E.~Mount, A.~O’Neil, K.~Shah, A.~K. Lee \emph{et~al.}, ``In
  silico labeling: Predicting fluorescent labels in unlabeled images,''
  \emph{Cell}, vol. 173, no.~3, pp. 792--803, 2018.

\bibitem{bastiaens1999fluorescence}
P.~I. Bastiaens and A.~Squire, ``Fluorescence lifetime imaging microscopy:
  spatial resolution of biochemical processes in the cell,'' \emph{Trends in
  cell biology}, vol.~9, no.~2, pp. 48--52, 1999.

\bibitem{wang2010image}
Q.~Wang, J.~Niemi, C.-M. Tan, L.~You, and M.~West, ``Image segmentation and
  dynamic lineage analysis in single-cell fluorescence microscopy,''
  \emph{Cytometry Part A: The Journal of the International Society for
  Advancement of Cytometry}, vol.~77, no.~1, pp. 101--110, 2010.

\bibitem{yuan2018computational}
H.~Yuan, L.~Cai, Z.~Wang, X.~Hu, S.~Zhang, and S.~Ji, ``Computational modeling
  of cellular structures using conditional deep generative networks,''
  \emph{Bioinformatics}, vol.~35, no.~12, pp. 2141--2149, 2018.

\bibitem{rivenson2019virtual}
Y.~Rivenson, H.~Wang, Z.~Wei, K.~de~Haan, Y.~Zhang, Y.~Wu, H.~G{\"u}nayd{\i}n,
  J.~E. Zuckerman, T.~Chong, A.~E. Sisk \emph{et~al.}, ``Virtual histological
  staining of unlabelled tissue-autofluorescence images via deep learning,''
  \emph{Nature biomedical engineering}, vol.~3, no.~6, p. 466, 2019.

\bibitem{lecun1998gradient}
Y.~LeCun, L.~Bottou, Y.~Bengio, and P.~Haffner, ``Gradient-based learning
  applied to document recognition,'' \emph{Proceedings of the IEEE}, vol.~86,
  no.~11, pp. 2278--2324, 1998.

\bibitem{simonyan2014very}
K.~Simonyan and A.~Zisserman, ``Very deep convolutional networks for
  large-scale image recognition,'' \emph{arXiv preprint arXiv:1409.1556}, 2014.

\bibitem{he2016deep}
K.~He, X.~Zhang, S.~Ren, and J.~Sun, ``Deep residual learning for image
  recognition,'' in \emph{Proceedings of the IEEE conference on computer vision
  and pattern recognition}, 2016, pp. 770--778.

\bibitem{szegedy2015going}
C.~Szegedy, W.~Liu, Y.~Jia, P.~Sermanet, S.~Reed, D.~Anguelov, D.~Erhan,
  V.~Vanhoucke, and A.~Rabinovich, ``Going deeper with convolutions,'' in
  \emph{Proceedings of the IEEE conference on computer vision and pattern
  recognition}, 2015, pp. 1--9.

\bibitem{wang2019deep}
H.~Wang, Y.~Rivenson, Y.~Jin, Z.~Wei, R.~Gao, H.~G{\"u}nayd{\i}n, L.~A.
  Bentolila, C.~Kural, and A.~Ozcan, ``Deep learning enables cross-modality
  super-resolution in fluorescence microscopy,'' \emph{Nature Methods},
  vol.~16, pp. 103--110, 2019.

\bibitem{weigert2018content}
M.~Weigert, U.~Schmidt, T.~Boothe, A.~M{\"u}ller, A.~Dibrov, A.~Jain,
  B.~Wilhelm, D.~Schmidt, C.~Broaddus, S.~Culley \emph{et~al.}, ``Content-aware
  image restoration: pushing the limits of fluorescence microscopy,''
  \emph{Nature methods}, vol.~15, no.~12, p. 1090, 2018.

\bibitem{Cai:KDD18}
L.~Cai, Z.~Wang, H.~Gao, D.~Shen, and S.~Ji, ``Deep adversarial learning for
  multi-modality missing data completion,'' in \emph{Proceedings of the 24th
  ACM SIGKDD International Conference on Knowledge Discovery and Data Mining},
  2018, pp. 1158--1166.

\bibitem{Zhang:NI15}
W.~Zhang, R.~Li, H.~Deng, L.~Wang, W.~Lin, S.~Ji, and D.~Shen, ``Deep
  convolutional neural networks for multi-modality isointense infant brain
  image segmentation,'' \emph{NeuroImage}, vol. 108, pp. 214--224, 2015.

\bibitem{LiMICCAI14}
R.~Li, W.~Zhang, H.-I. Suk, L.~Wang, J.~Li, D.~Shen, and S.~Ji, ``Deep learning
  based imaging data completion for improved brain disease diagnosis,'' in
  \emph{Proceedings of the 17th International Conference on Medical Image
  Computing and Computer Assisted Intervention}, 2014, pp. 305--312.

\bibitem{Chen:KDD18}
Y.~Chen, H.~Gao, L.~Cai, M.~Shi, D.~Shen, and S.~Ji, ``Voxel deconvolutional
  networks for {3D} brain image labeling,'' in \emph{Proceedings of the 24th
  ACM SIGKDD International Conference on Knowledge Discovery and Data Mining},
  2018, pp. 1226--1234.

\bibitem{vaswani2017attention}
A.~Vaswani, N.~Shazeer, N.~Parmar, J.~Uszkoreit, L.~Jones, A.~N. Gomez,
  {\L}.~Kaiser, and I.~Polosukhin, ``Attention is all you need,'' in
  \emph{Advances in Neural Information Processing Systems}, 2017, pp.
  6000--6010.

\bibitem{wang2018global}
Z.~Wang, N.~Zou, D.~Shen, and S.~Ji, ``Global deep learning methods for
  multimodality isointense infant brain image segmentation,'' \emph{arXiv
  preprint arXiv:1812.04103}, 2018.

\bibitem{kolda2009tensor}
T.~G. Kolda and B.~W. Bader, ``Tensor decompositions and applications,''
  \emph{SIAM review}, vol.~51, no.~3, pp. 455--500, 2009.

\bibitem{ronneberger2015u}
O.~Ronneberger, P.~Fischer, and T.~Brox, ``U-net: Convolutional networks for
  biomedical image segmentation,'' in \emph{International Conference on Medical
  image computing and computer-assisted intervention}.\hskip 1em plus 0.5em
  minus 0.4em\relax Springer, 2015, pp. 234--241.

\bibitem{huang2017densely}
G.~Huang, Z.~Liu, L.~Van Der~Maaten, and K.~Q. Weinberger, ``Densely connected
  convolutional networks.'' in \emph{CVPR}, vol.~1, no.~2, 2017, p.~3.

\bibitem{quan2016fusionnet}
T.~M. Quan, D.~G. Hildebrand, and W.-K. Jeong, ``Fusionnet: A deep fully
  residual convolutional neural network for image segmentation in
  connectomics,'' \emph{arXiv preprint arXiv:1612.05360}, 2016.

\bibitem{fakhry2017residual}
A.~Fakhry, T.~Zeng, and S.~Ji, ``Residual deconvolutional networks for brain
  electron microscopy image segmentation,'' \emph{IEEE transactions on medical
  imaging}, vol.~36, no.~2, pp. 447--456, 2017.

\bibitem{ioffe2015batch}
S.~Ioffe and C.~Szegedy, ``Batch normalization: Accelerating deep network
  training by reducing internal covariate shift,'' in \emph{International
  Conference on Machine Learning}, 2015, pp. 448--456.

\bibitem{srivastava2014dropout}
N.~Srivastava, G.~Hinton, A.~Krizhevsky, I.~Sutskever, and R.~Salakhutdinov,
  ``Dropout: a simple way to prevent neural networks from overfitting,''
  \emph{The Journal of Machine Learning Research}, vol.~15, no.~1, pp.
  1929--1958, 2014.

\bibitem{kingma2014adam}
D.~P. Kingma and J.~Ba, ``Adam: A method for stochastic optimization,''
  \emph{arXiv preprint arXiv:1412.6980}, 2014.

\end{thebibliography}
\bibliographystyle{IEEEtran}

\end{document}